\documentclass[12pt,twocolumn]{iopart}

\usepackage{graphicx}
\usepackage[usenames]{color}
\newcommand{\beq}{\begin{equation}}
\newcommand{\eeq}{\end{equation}}
\newcommand{\beqa}{\begin{eqnarray}}
\newcommand{\eeqa}{\end{eqnarray}}
\newcommand{\ba}{\begin{array}}
\newcommand{\ea}{\end{array}}

\begin{document}
\title{Localization of a dipolar Bose-Einstein condensate
in a bichromatic optical lattice}

\author{P. Muruganandam$^{1,2}$\footnote{murganand@gmail.com},
R. Kishor Kumar$^2$, and S. K. Adhikari$^1$
 \footnote{adhikari@ift.unesp.br; URL: www.ift.unesp.br/users/adhikari}}
\address{$^1$Instituto de F\'{\i}sica Te\'orica, UNESP - Universidade
Estadual Paulista,\\
01.140-070 S\~ao Paulo, S\~ao Paulo, Brazil\\
$^2$School of Physics, Bharathidasan University, Palkalaiperur Campus,\\
Tiruchirappalli  620024, Tamilnadu, India}

\begin{abstract} By numerical simulation and variational analysis of the 
Gross-Pitaevskii equation we study the localization, with an exponential 
tail, of a dipolar Bose-Einstein condensate (DBEC) of $^{52}$Cr atoms in 
a three-dimensional bichromatic optical-lattice (OL) generated by two 
monochromatic OL of incommensurate wavelengths along three orthogonal 
directions. For a fixed dipole-dipole interaction, a localized state of 
a small number of atoms ($\sim 1000$) could be obtained when the
 short-range interaction is not too attractive or not too repulsive. A 
phase diagram showing the region of stability of a DBEC with short-range 
interaction and dipole-dipole interaction is given.
 \end{abstract}


\maketitle

\section{Introduction}The localization of a non-interacting wave form in a
 disordered potential, predicted by Anderson \cite{anderson}, has
been the topic of vigorous research in different areas. Localization has been
observed experimentally in diverse contexts \cite{chabe}
including
Bose-Einstein condensates (BEC) \cite{billy,roati}.
Billy {\it et al.} \cite{billy} demonstrated the
localization of a cigar-shaped interacting $^{87}$Rb BEC released
into a one-dimensional
(1D) waveguide with controlled disorder created by a laser speckle.
Roati {\it et al.} \cite{roati} observed  the localization of a
 non-interacting $^{39}$K BEC in a bichromatic
quasi-periodic optical-lattice (OL) potential created by
superposing of two standing-wave polarized laser beams with incommensurate
wavelengths.
The non-interacting BEC was created \cite{roati} by
tuning the atomic scattering length $a$ to zero near a Feshbach
resonance \cite{fesh}. 
The disorder   in a quasi-periodic OL potential is
deterministic, in contrast to the complete disorder in a optical speckle
potential. The  localization in such a  quasi-periodic  potential
  is a special case of Anderson
localization in a fully disordered potential
 and is well described by
the Aubry-Andr\'e model \cite{aubry}. Such a localization is often termed
 Aubry-Andr\'e localization. However, these two mechanisms of localization are 
distinct. 
While Anderson localization of wave functions with
exponential tails is a pure quantum effect, the Aubry-Andr\'e 
localization may occur in a classical phase
space \cite{Albert}.
The localization of a BEC in a disordered
potential has been the subject matter of various theoretical
\cite{boers,modugno,adhikari,cheng,adhikari1}
and experimental \cite{chabe,billy,roati} studies.

In the presence of strong disorder one has strong Anderson localization 
\cite{billy,roati}, where the localized state could be quite similar to 
a localized state of Gaussian shape in an infinite potential or a 
potential of very high barriers. Then the quantum state cannot escape 
the strong barriers of the disordered potential. However, the more 
interesting case of localization is in the presence of a weak disorder 
when the system is localized due to the quasi-periodic (disordered) 
nature of the potential \cite{billy,roati} and not due to the strength 
of the lattice. The localization takes place due to cancellation of 
waves coming after multiple scattering from many barriers of the random 
potential. When this happens the localized state acquires a pronounced 
exponential tail.

The usual dilute BEC with negligible dipole moment
interacting via short-range
interaction 
is described by the mean-field
Gross-Pitaevskii (GP) equation. More recently, it has been
possible \cite{pfau}
to obtain
a BEC of $^{52}$Cr atoms with large dipole moment leading to a long-range
dipole-dipole  (dipolar) interaction  superposed on the usual
short-range interaction, which can be varied using  a Feshbach
resonance \cite{fesh}. This allows  to study the
dipolar BEC (DBEC)
of $^{52}$Cr atoms with variable  short-range interaction \cite{pfau}.
Because of the anisotropic dipolar interaction, the DBEC possesses many
distinct features \cite{pfau,jb,Dutta2007,Parker2009,other},
which are under active investigation by  different research groups
\cite{metz,coll}. For example, the stability of a DBEC depends not only 
on the value of the scattering length, but also strongly on the geometry 
of the trapping potential \cite{pfau,jb,Dutta2007}. A
pancake-shaped trapping potential
give a repulsive mean field of dipolar interaction 
 and thus the dipolar condensate is
more stable. In contrast, a cigar-shaped trapping potential 
yields an attractive mean-field of dipolar interaction
 and hence leading to a dipolar
collapse
\cite{pfau,Ronen2006a,Yi2000,Yi2003,Santos2000}.
Peculiarities in the collective low energy shape
oscillations of DBEC have been studied by Yi and You \cite{Yi2000}.
Analogue of roton-maxon instability and the  appearance
of roton minimum in Bogoliubov spectrum \cite{Wilson2008,Ronen2007} and  
angular collapse of DBEC 
\cite{Wilson2009} have been studied.
Numerical studies on the
stability shows certain unusual structure of the Hartree ground state of
dipolar condensate in an anisotropic trap~\cite{Dutta2007,Parker2009}.

After the experiment \cite{roati} on the localization of a 1D BEC in a 
quasi-periodic trap, a natural extension of this phenomenon would be to 
achieve localization in two and three dimensions (3D) 
\cite{2D3D,adhikari1}, both for a BEC and a DBEC. The theoretical 
description of a dilute weakly-interacting DBEC can be formulated by 
including a dipolar interaction term in the GP equation 
maintaining the formal simplicity, nevertheless, increasing vastly the 
numerical complexity. Using the numerical and variational solutions of 
the GP equation, we study the localization of a BEC and a DBEC in 3D, in 
the presence of bichromatic OL potentials along orthogonal directions. 
Although the localized states have an exponential tail in a weak 
quasi-periodic potential, the central part of such localized states, 
responsible for the major contribution to the total density, may have a 
Gaussian distribution. In the present paper we shall consider such 
localized states with an exponential tail and a Gaussian central part, 
which can also be studied by the variational approximation. The 
variational approximation provides an analytical understanding of the 
localization and also yields interesting result when the numerical 
procedure is difficult to implement.  Such a variational Gaussian 
approximation has successfully been applied to the localization of a BEC 
without dipolar interaction in one \cite{adhikari}, 
two and three dimensions 
\cite{adhikari1} in a 
bichromatic OL potential as well as a speckle potential \cite{cheng} in one 
dimension.

Due to the
angle dependence of the long-range 
dipolar interaction, the localization of a DBEC is
more interesting than a BEC with only a short-range interaction. 
The dipolar interaction is weak  in the
spherically-symmetric shape, is attractive
in  the cigar shape (aligned
dipoles arranged linearly attract each other) and  repulsive
 in the  pancake shape (aligned dipoles arranged
side-by-side repeal each other) \cite{coll}. Because of this,
 the trap configuration
plays an essential role in the localization of a DBEC. The controllable
short-range interaction together with the exotic dipolar interaction makes the DBEC an attractive system for
experimental localization in a bichromatic OL potential
and a challenging system for
theoretical investigation allowing to study the interplay between the
dipolar interaction and the short-range interaction in dipolar atoms.

The cigar- and pancake-shaped localized DBEC are obtained by considering
bichromatic OL potentials of different strengths along the orthogonal
directions. A localized DBEC, without short-range interaction, 
can be achieved in a
bichromatic OL trap for a small number of atoms by tuning \cite{pfau}
the atomic scattering length to zero near a Feshbach resonance
\cite{fesh}. For a cigar-shaped DBEC, without short-range interaction, as the number of
atoms increases it becomes highly attractive and suffers from collapse
instability. In the presence of a repulsive short-range interaction, the effect of the
attractive dipolar interaction in the cigar shape can be compensated leading to a
localized DBEC for a small number of atoms; delocalization may take
place due to excess of repulsion for a large number of atoms. For a
pancake-shaped DBEC, the dipolar interaction is repulsive and a localized DBEC is
obtained for a small number of atoms. For a large number of atoms excess
of repulsion should lead to delocalization.  From a variational analysis
of the localization of a DBEC, a phase diagram illustrating its
stability for different short-range interaction, number of atoms, and the geometry of the
bichromatic OL is given.  To obtain localization, the short-range interaction should be
small. As in 1D \cite{modugno,adhikari,destruction}, a large repulsive
short-range interaction destroys the localization and the localized state escapes to
infinity in all cases. A large attractive short-range interaction leads to collapse
instability and destroys the localization.

In Sec. II we present a brief account of the modified GP equation with
the bichromatic OL potential in the presence of a dipolar interaction together with a
Gaussian variational analysis. In Sec. III we present numerical
and variational studies  of localization of a dipolar dipolar interaction in the
presence and absence
of a short-range interaction. In Sec. IV we present a brief summary of  the present
investigation.

\section{Analytical Formulation} We shall study the localization of the 
DBEC of $N$ atoms, each of mass $m$, using the following mean-field GP
equation with bichromatic OL potential $V({\bf r})$: \cite{pfau}
\begin{eqnarray}  \label{gp3d}\fl  \frac{m}{\hbar^2}\mu \phi({\bf r})
 = \biggr[ -\frac{1}{2}\nabla^2 +V({\bf r}) + 4\pi a N|\phi({\bf r})|^2
+ \int U_{dd}({\bf r -r'})|\phi({\bf r'})|^2d{\bf r'}
\biggr] \phi({\bf r}), \end{eqnarray} 
with 
\begin{eqnarray}
 U_{dd}({\bf r}) = 3N
a_{dd}(1-3\cos^2\theta)/r^3,
\end{eqnarray}
 and \begin{eqnarray}\label{pot}
 V({\bf r}) = \sum_{l=1}^2 A_l[
\nu_\rho \{ \sin ^2\left(k_l x\right)+ \sin ^2\left(k_l y\right)\}
+\nu_z \sin ^2\left(k_l z\right)],
\end{eqnarray}
 where $\phi({\bf r})$ is the DBEC
wave function of normalization $\int \phi({\bf r})^2 d {\bf r}=1$, $\mu$
is the chemical potential, $a$ the atomic scattering length, $\theta$ is
the angle between $\bf r$ and the direction of polarization, here taken
along the
 $z$ axis, $A_l=k_l^2s_l/2$, $\lambda_l$'s are the wave lengths,
 $k_l=2\pi/\lambda_l$ are the wave numbers, and $s_l$ are the strengths
of the OL potentials.  The parameters $\nu_\rho$ and $\nu_z$ control the
relative strengths of the OL's in different directions and can be varied
to achieve the pancake- ($\nu_z>1, \nu_\rho=1$) and cigar-shaped
($\nu_\rho>1, \nu_z=1$) DBEC. The constant $a_{dd}
=\mu_0\bar \mu^2 m /(12\pi \hbar^2)$ in the dipolar interaction $U_{dd}$
is a length characterizing the strength of 
dipolar interaction and its experimental
value for $^{52}$Cr atoms is $15a_0$ \cite{pfau}, with  $a_0$ the Bohr 
radius, 
 $\bar \mu$ the (magnetic) dipole moment of a single atom, and $\mu_0$ 
the permeability of free space.

The GP equation (\ref{gp3d}) can be solved variationally by minimizing 
the energy functional \cite{pfau} \begin{eqnarray}\label{ene}\fl
\frac{m}{\hbar^2}E[\phi]= \int\biggr[ \frac{1}{2} \vert\nabla
\phi\vert^2 + \sum_{j=1}^d V({\bf r})\phi^2+2\pi a N \phi^4 
 + \frac{\phi^2}{2}\int U_{dd}({\bf r-r'})|\phi({\bf r '})|^2d{\bf
r'} \biggr]d{\bf r} \end{eqnarray} with the following Gaussian ansatz for
$\phi(\bf r)$ \cite{you}: \begin{equation}\label{ans} \phi({\bf
r})=\left(\frac{\pi^{-3/2}}{w_\rho^2 w_z}\right)^{1/2} \exp\left(-
\frac{\rho^2}{2w_\rho^2}- \frac{z^2}{2w_z^2} \right), \end{equation}
where $w_\rho$ is the width in the radial direction $\rho$ and $w_z$ is
the width in the axial direction $z$. We have assumed circular symmetry
in the $x-y$ plane. Equations (\ref{ene}) and (\ref{ans}) lead to
\cite{you} \begin{eqnarray} \label{enmin}
 \frac{m}{\hbar^2}E=& \, 
\frac{1}{4}\left[\frac{2}{w_\rho^2}+\frac{1}{w_z^2} \right] +\frac{N
}{\sqrt{2\pi}} \frac{1}{w_\rho^2 w_z}\left[a -a_{dd}f(\kappa)
\right]\nonumber \\ & \, + \frac{1}{2}\sum_{l=1}^2 A_l
\left[2\nu_\rho+\nu_z- 2\nu_\rho{\cal E}_{\rho,l} -\nu_z{\cal
E}_{z,l}\right], \\ f(\kappa)=& \,
\frac{1+2\kappa^2}{1-\kappa^2}-\frac{3\kappa^2
\mbox{atanh}\sqrt{1-\kappa^2}} {(1-\kappa^2)^{3/2}}, \quad \kappa\equiv
\frac{w_\rho}{w_z}, \label{kappa} \end{eqnarray} where $ {\cal E}_{\rho,
l}= \exp(-w_\rho^2k_l^2 ),
 {\cal E}_{z,l}= \exp(-w_z^2k_l^2 ). $ For a stationary state the energy
$mE/\hbar^2$ should have a minimum as a function of $w_\rho$ and $w_z$:
$\partial E/\partial w_\rho=\partial E/\partial w_z=0, $ together with
the condition 
\begin{equation} \left(\frac{\partial^2 E}{\partial
w_\rho\partial w_z}\right )^2- \frac{\partial^2 E}{\partial w_\rho^2}
\frac{\partial^2 E}{\partial w_z^2}<0. \end{equation}
 The minima conditions $\partial E/\partial w_\rho=\partial E/\partial
w_z=0 $ become in explicit notation \begin{eqnarray}  \label{e1} && 2\nu_\rho
w_\rho^4\sum_{l=1}^2A_lk_l^2{\cal E}_{\rho,l} -\frac{N}{\sqrt{2\pi}
w_z}\left[{2a} -a_{dd}g(\kappa)\right] =1,\\  \label{e2}&&
2\nu_zw_z^4\sum_{l=1}^2 A_lk_l^2{\cal E}_{z,l}
-\frac{2Nw_z}{\sqrt{2\pi}w_\rho^2}\left[{a}-a_{dd} {h(\kappa)}\right]
=1,\\ \mbox{where} \nonumber \\ 
\label{e3} &&
g(\kappa)=\frac{2-7\kappa^2-4\kappa^4}{(1-\kappa^2)^2}+
\frac{9\kappa^4\mbox{atanh}\sqrt{1-\kappa^2}}{(1-\kappa^2)^{5/2}},\\
\label{e4} &&
h(\kappa)=\frac{1+10\kappa^2-2\kappa^4}{(1-\kappa^2)^2}
-\frac{9\kappa^2\mbox{atanh}\sqrt{1-\kappa^2}}{(1-\kappa^2)^{5/2}}.
\end{eqnarray}
 The solution of  (\ref{e1}) $-$ (\ref{e4}) determine the widths
$w_z$ and $w_\rho$. For a certain set of parameters, localization is
possible if there is a minimum of the energy. Hence from 
(\ref{enmin}) and (\ref{e1}) $-$ (\ref{e4}) we can determine if a BEC or
DBEC would be localized or not.

\section{Numerical Study}
We perform a full 3D numerical simulation in Cartesian $x,y,z$
variables using imaginary- and real-time propagation with
Crank-Nicolson discretization \cite{murug}
employing  small space
($\sim 0.025$)
and time ($\sim 0.0001$) steps necessary for
obtaining converged results.  For this purpose
we use the
FORTRAN programs provided in  \cite{CPC} after transforming 
(\ref{gp3d}) into a time-dependent form by replacing $m\mu/\hbar^2$ by
$i\partial /\partial t$, where $t$ is time.  The dipolar interaction
 term is
evaluated by the usual fast Fourier transformation technique \cite{jb}.
The imaginary- and real-time propagation lead essentially to the same
localized states. This not only assures the correctness of the calculational
scheme, but also guarantees that the localized states are stationary and not
a result of dynamical self-trapping \cite{self}. 
(The imaginary-time propagation can
only find the stationary localized states and
cannot obtain the dynamical self-trapped states.)
The stability of the localized state was tested by
real-time propagation allowing  small perturbations of potential or interaction
parameters.  (In the absence of the dipolar interaction, the localized states have been
demonstrated explicitly to be stable  in one \cite{adhikari}
and two \cite{adhikari1} dimensions.)
The accuracy of the
numerical simulation was tested by varying the size of
space and time steps and the total number of space and time steps.

\begin{figure}
\begin{center}
\includegraphics[width=.49\linewidth]{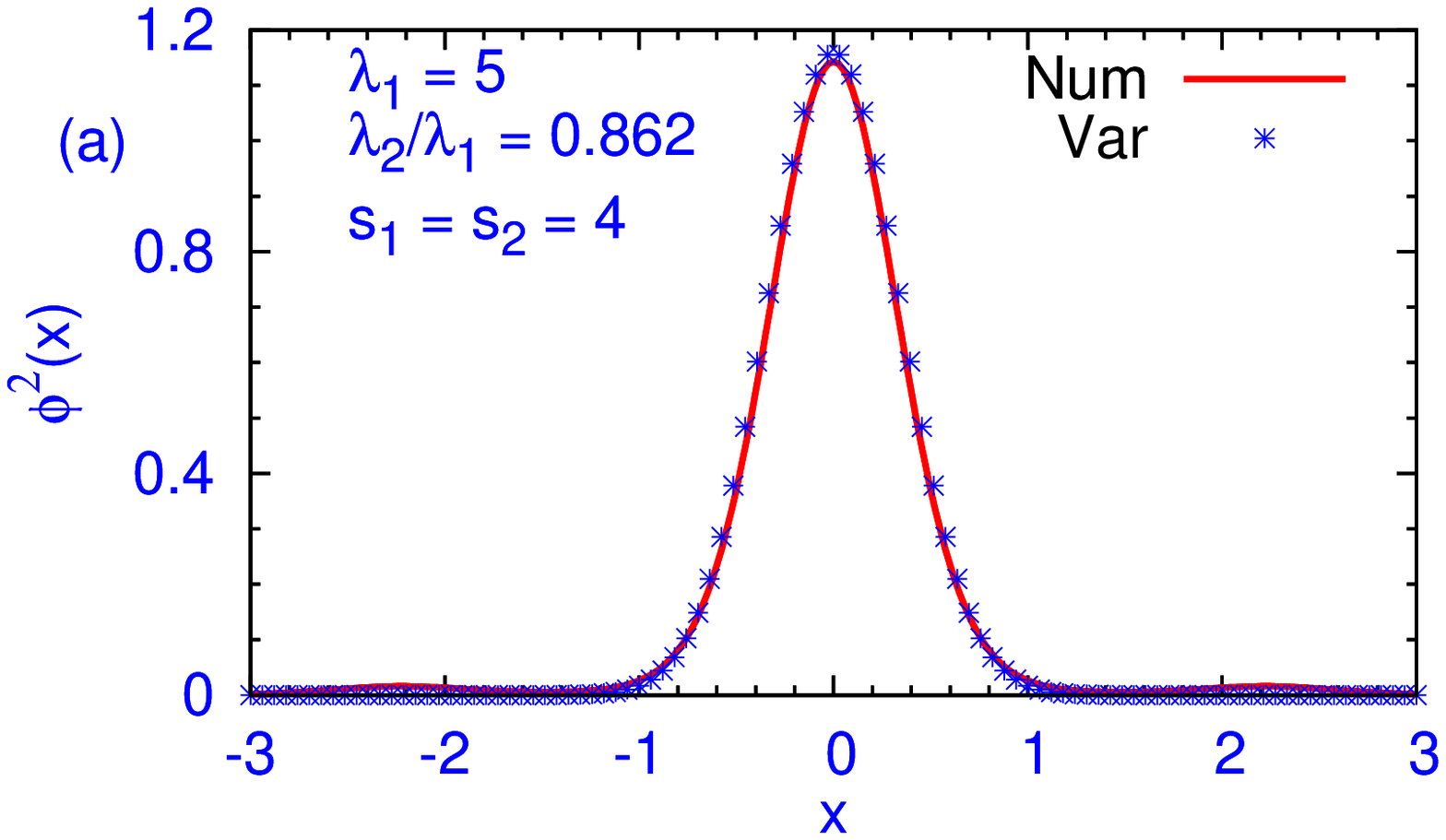}
\includegraphics[width=.49\linewidth]{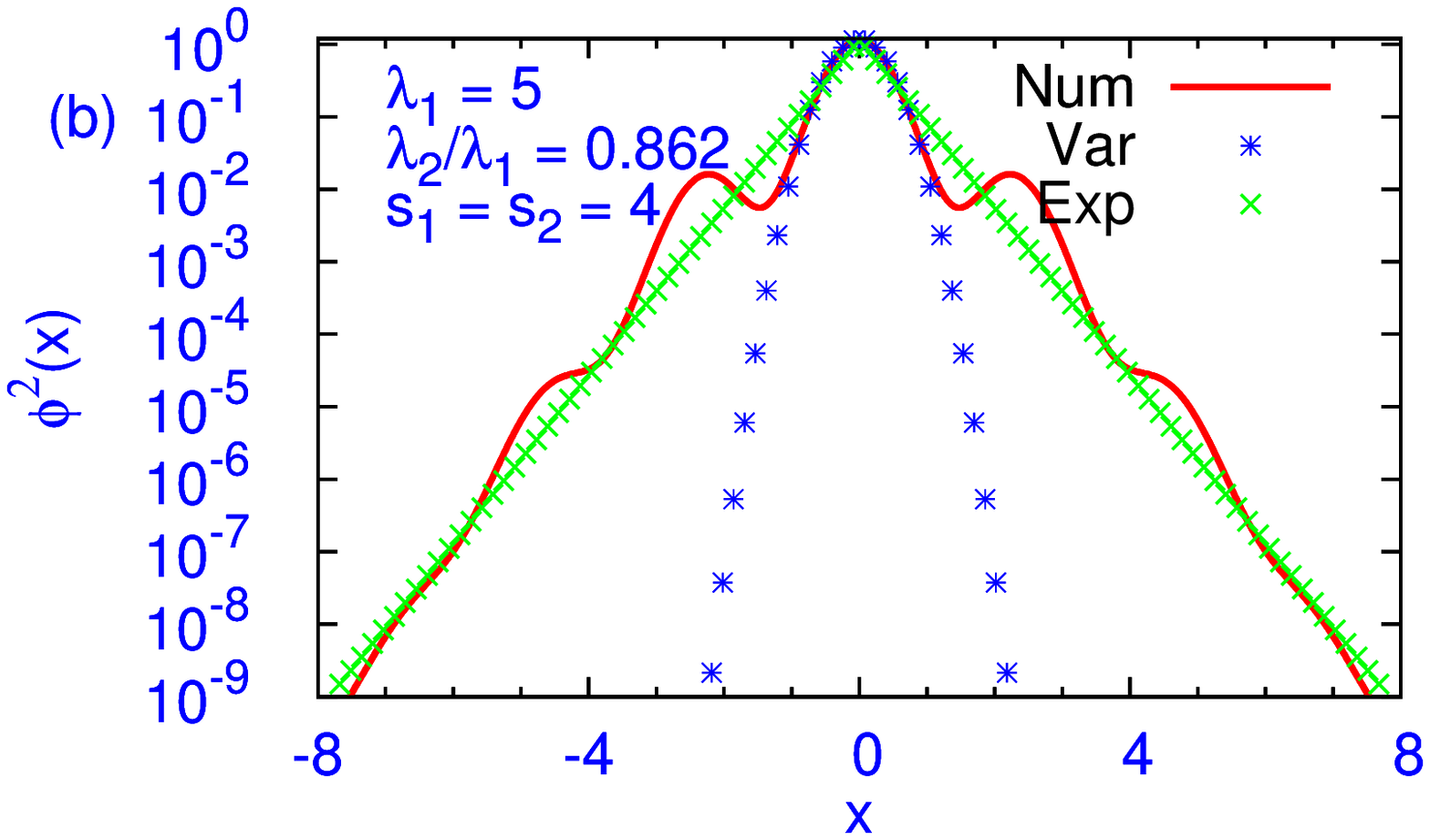}
\end{center}

\caption{(Color online) (a) Normalized numerical (Num) and variational
(Var)   densities
$\phi^2 (x)$ of  (\ref{1D})
versus  $x$ for the 1D localized state with bichromatic OL wave lengths 
$\lambda_1 = 5,
\lambda_2 =0.862 \lambda_1,$ and strengths $ s_1=s_2=4$.
(b) The same densities together
with the
exponential fit $\phi^2 (x)=\exp[-2\mathrm{abs}(x)/l_{\mathrm{loc}}]/
(0.4865\sqrt\pi),
l_{\mathrm{loc}}=0.75$
on a log scale.
}\label{fig1} \end{figure}

\begin{figure}
\begin{center}
\includegraphics[width=.49\linewidth]{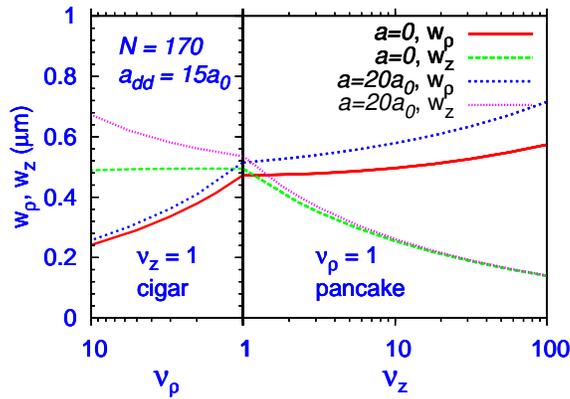}
\end{center}

\caption{(Color online)  Variational widths $w_\rho$ and $w_z$  of the
DBEC versus bichromatic OL relative strength 
parameters $\nu_\rho$ and $\nu_z$  [viz. Eq. (\ref{pot})] for 
170 $^{52}$Cr atoms
for scattering length $a=0$ and $20a_0$.}
\label{fig2} \end{figure}

To compare the numerical results with variational analysis, we only
consider localized states mostly occupying a single site of the
bichromatic OL.   Throughout this study the strength parameters of the
two components 
of the 
 bichromatic OL are taken as $s_1=s_2=4$ and the corresponding wave lengths 
are taken as 
$\lambda_1=5$
$\mu$m, $\lambda_2=0.862\lambda_1.$ 
The relative strength of the radial  and  axial bichromatic optical 
lattice are varied by adjusting the parameters $\nu_\rho$ and 
$\nu_z$ in Eq. (\ref{pot}).
To demonstrate that with these sets
of parameters we are in the limit of   
localization with an exponential tail in a weak potential, we
solve the 1D linear Schr\"odinger equation in dimensionless variables
\cite{CPC} \begin{equation}\label{1D} -\frac{1}{2}\phi
''(x)+\sum_{l=1}^2 \frac{2\pi^2s_l} {\lambda_l^2} \sin^2
\left(\frac{2\pi}{\lambda_l}x\right) \phi(x)={\cal E } \phi(x),
\end{equation} with the above sets of parameter,
 where the prime denotes $x$-derivative and $\cal E$ denotes energy. In
the limit of zero nonlinearity ($a=0$) 
and zero dipolar interaction ($a_{dd}=0$), 
(\ref{gp3d}) decouples into three equations like  (\ref{1D}). In figure
\ref{fig1} (a) and (b) we plot the density $\phi^2(x)$ of  (\ref{1D})
versus $x$ in linear and log scales together with its Gaussian
variational counterpart $\phi^2(x)=\exp(-x^2/w^2)/(\pi^{1/2}w)$, with
$w=0.4865$ the variational width, and the exponential fit $\phi^2(x)=
\exp [-{\mathrm{abs}}(x)/ l_{\mathrm{loc}}] /(0.4865\pi^{1/2}) $, with
$l_{\mathrm{loc}}=0.75$ the localization length \cite{billy} providing a
measure of the exponential tail. For strong  localization with
strong bichromatic potential $l_{\mathrm{loc}}\to 0$ and the
localized state has a pure Gaussian tail. However, when $
l_{\mathrm{loc}} > x_{rms}$ with $x_{rms}$ the root mean square size,
the exponential tail is pronounced and the limit of 
localization in a weak quasi-periodic potential 
is attained \cite{billy,roati}. In the present example,
$x_{rms}=0.53$ and $l_{\mathrm{loc}}=0.75$, and hence we are in the
limit of localization in a weak potential. This is clear from  figure
 \ref{fig1}
(b), where the central part of the localized state is fitted to the
variational Gaussian solution, whereas the large-$x$ parts are fitted to
an exponential function 
with a large localization length over about nine orders
of magnitude. In  figure
 \ref{fig1} (b), one can identify several minor
peaks in successive wells of the bichromatic OL potential. The state of
 figure
 \ref{fig1} is  localized by the weak bichromatic lattice and if we
substantially reduce the strength of the potential, no localized state
will emerge. In the experiments of Refs. \cite{billy,roati} and in some
other studies on  localization \cite{modugno} in a weak potential, localized
states with a pronounced undulating tail over many wells of the
localization potential were considered. However, a weak limit of 
localization with an exponential tail 
can also be achieved in the absence of a pronounced
undulating tail, as we have shown here. In this paper we shall be
considering localization in the presence of a pronounced
Gaussian peak and a weak undulating tail as can be seen from  figure
\ref{fig1} (b). The existence of a pronounced Gaussian peak will be
turned to a good advantage in predicting accurate analytical variational
results for the localized state and for an analytical understanding of
the localization. We shall also consider the  localization in
the presence of short-range interaction and dipolar interaction. The inclusion of a repulsive 
short-range interaction will in
general destroy localization by increasing the localization length and
thus creating a more pronounced exponential tail \cite{modugno}. The
inclusion of the dipolar interaction will have an effect on localization, which we shall
study here, and such inclusion should not destroy the exponential tail
of  localization as illustrated in  figure
 \ref{fig1} (b).

\begin{figure}
\begin{center}
\includegraphics[width=.49\linewidth]{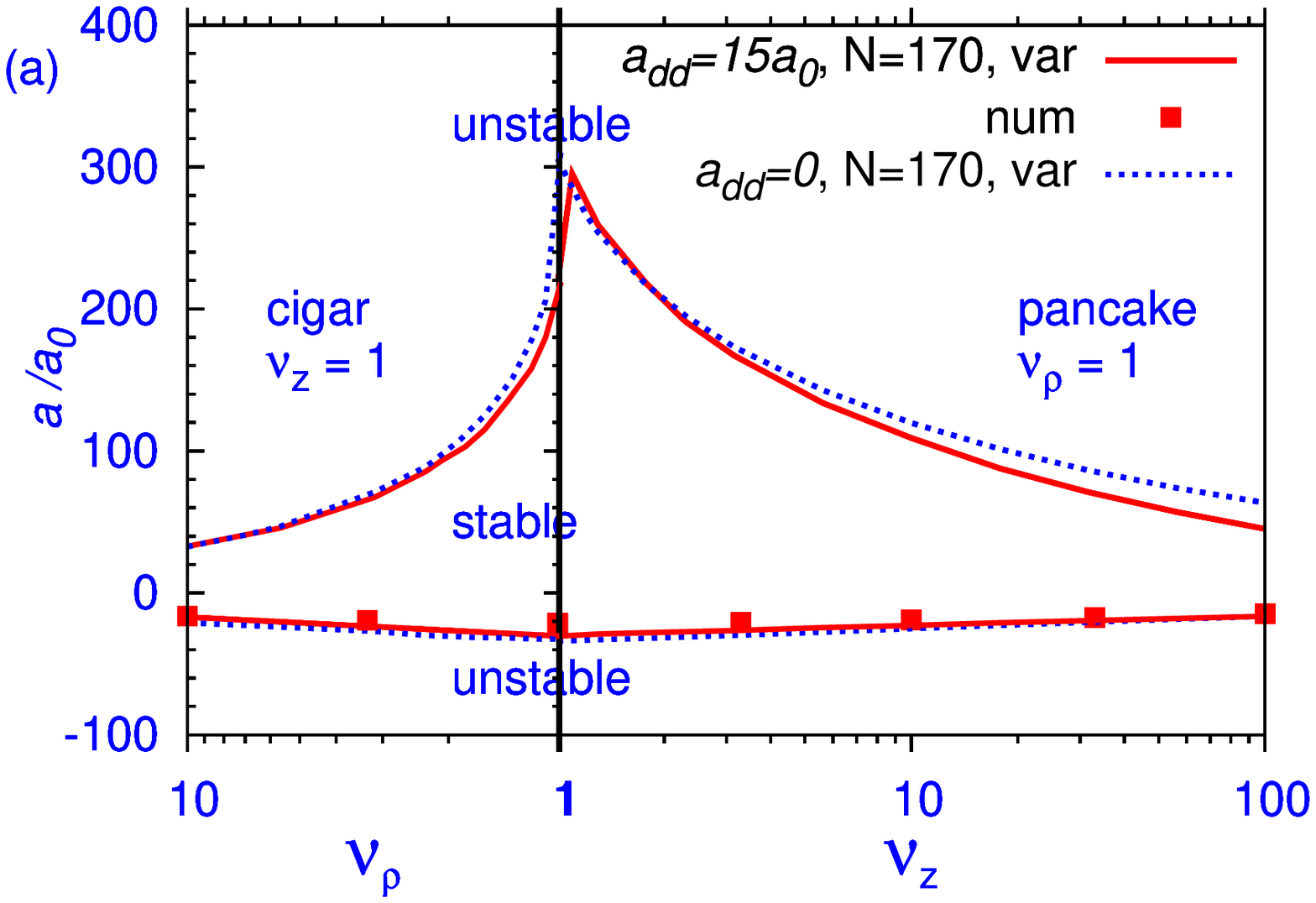}
\includegraphics[width=.49\linewidth]{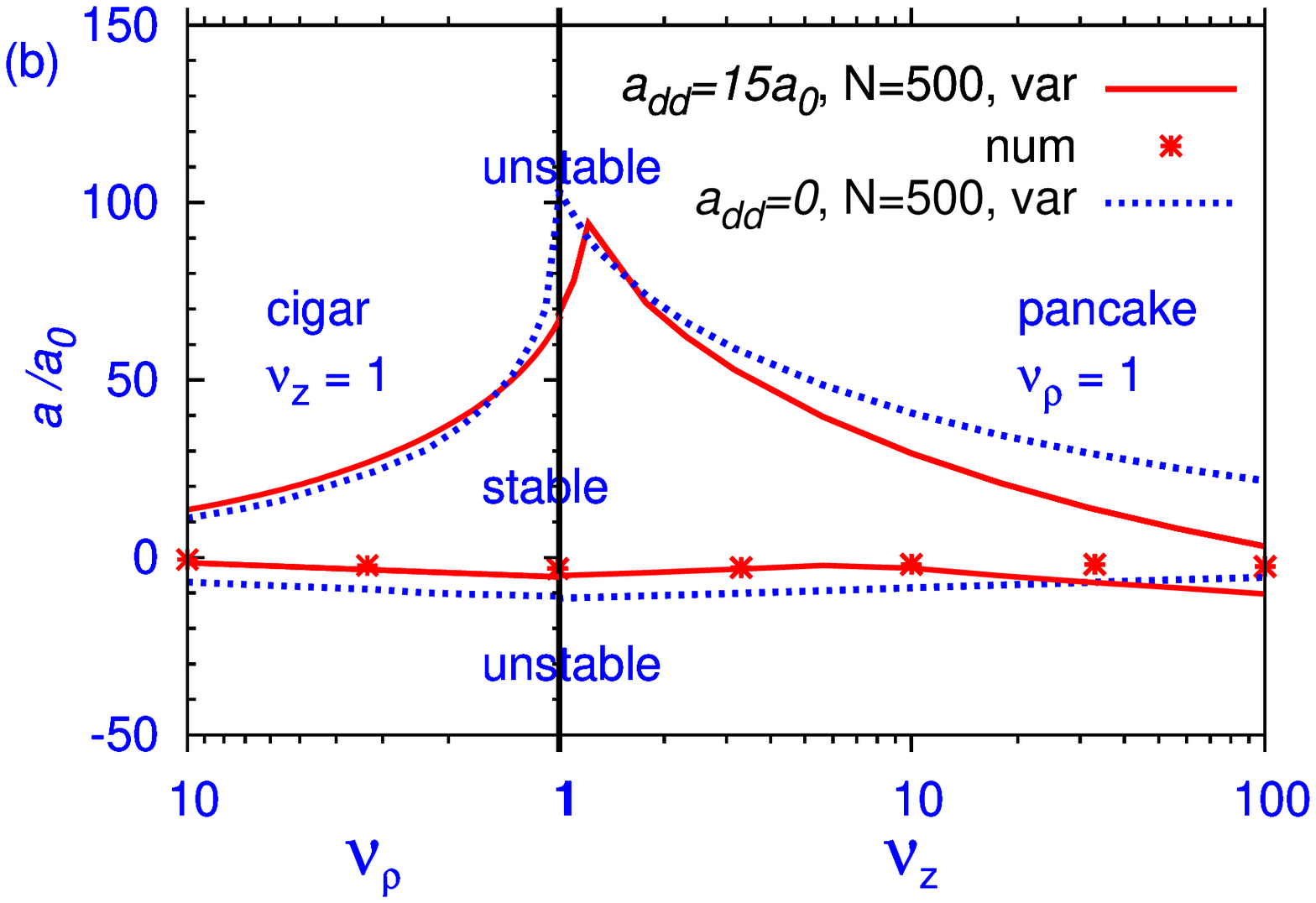}
\end{center}

\caption{(Color online) 
 Stability region in the $a/a_0$ versus $\nu_\rho$ and $\nu_z$ (relative 
strengths of the bichromatic OL in radial and axial directions)
phase
plots for  (a)  170  and (b) 500 atoms for $a_{dd}=15a_0$ ($^{52}$Cr atoms) and $a_{dd}=0$
variational - var (lines), numerical - num (points).
Localization is possible between the upper
and lower lines of the same data set.}
\label{Fig2} \end{figure}

A variational
analysis is useful for a qualitative understanding of the problem and we
present the same before considering a numerical solution of 
(\ref{gp3d}).
In  figure
 \ref{fig2}  we plot the variational widths
$w_\rho$ and $w_z$ for a DBEC of 170 $^{52}$Cr atoms in the cigar and
pancake shapes. This figure shows the evolution of the widths from the
pancake to cigar shapes while the radial width $w_\rho$ reduces and the 
axial width $w_z$ increases as expected. For $a=0$, the axial width $w_z$ 
does not increase with the increase of $\nu_\rho$, as in the cigar shape 
the  dipolar interaction becomes attractive and does not permit the increase of $w_z$.  

Next we study, using  (\ref{enmin}), the set
of values of the parameters for which the energy can have a minimum and
allows a stable localized DBEC. The region of stability for dipole
strength $a_{dd}=0$ and $=15a_0$ is shown in  figure
\ref{Fig2} 
for (a) 170 and (b) 500 $^{52}$Cr
atoms as a phase  plot of $a/a_0$ versus the relative strengths of the 
bichromatic OL 
$\nu_\rho$
and $\nu_z$ in radial and axial directions. The stability
of a DBEC in a harmonic trap has also been studied \cite{jb}.
For a fixed $a_{dd}$ and $N$ the stability region lies
between the two corresponding lines in  figure
 \ref{Fig2}.
Above the upper line the system
becomes too repulsive (positive $Na/a_0$) to be confined by the
weak bichromatic OL. Below the lower line the system becomes too attractive
(negative $Na/a_0$) and suffers from collapse instability. The
localization of a BEC (without dipolar interaction) is controlled by the
non-linearity $4\pi a N$ alone of  (\ref{gp3d})
and the effect of
the dipolar interaction on the 
stability of the DBEC is clearly exhibited in  figures
\ref{Fig2}. In these figures we also plot numerical results for
collapse instability with negative (attractive) scattering length 
(the lower limit of stability in this figure),
which agree well with the variational results. Starting from a stable localized 
state, 
the stability lines (points) are 
obtained by slowly changing the scattering length $a$ at a fixed trap symmetry 
(by fixing the parameters $\nu_\rho$ and $\nu_z$) until no localized state 
can be obtained (by numerical or variational means).
In the pancake side the
dipolar interaction is repulsive and the localization is 
destroyed for a smaller
value of the repulsive scattering length compared to the case where
dipolar interaction is absent, as can be seen in  figure
 \ref{Fig2}.
 The opposite happens in the cigar side where the dipolar 
interaction is attractive. In the cigar side the dipolar 
interaction is attractive and the localization is destroyed for a larger 
value of the repulsive scattering length compared to the case where 
dipolar interaction is absent, as can be seen in figure
 \ref{Fig2} (b) for 500 atoms. This effect is  smaller in figure
 \ref{Fig2} (a)  for 170 atoms.

\begin{figure}
\begin{center}
\includegraphics[width=.49\linewidth]{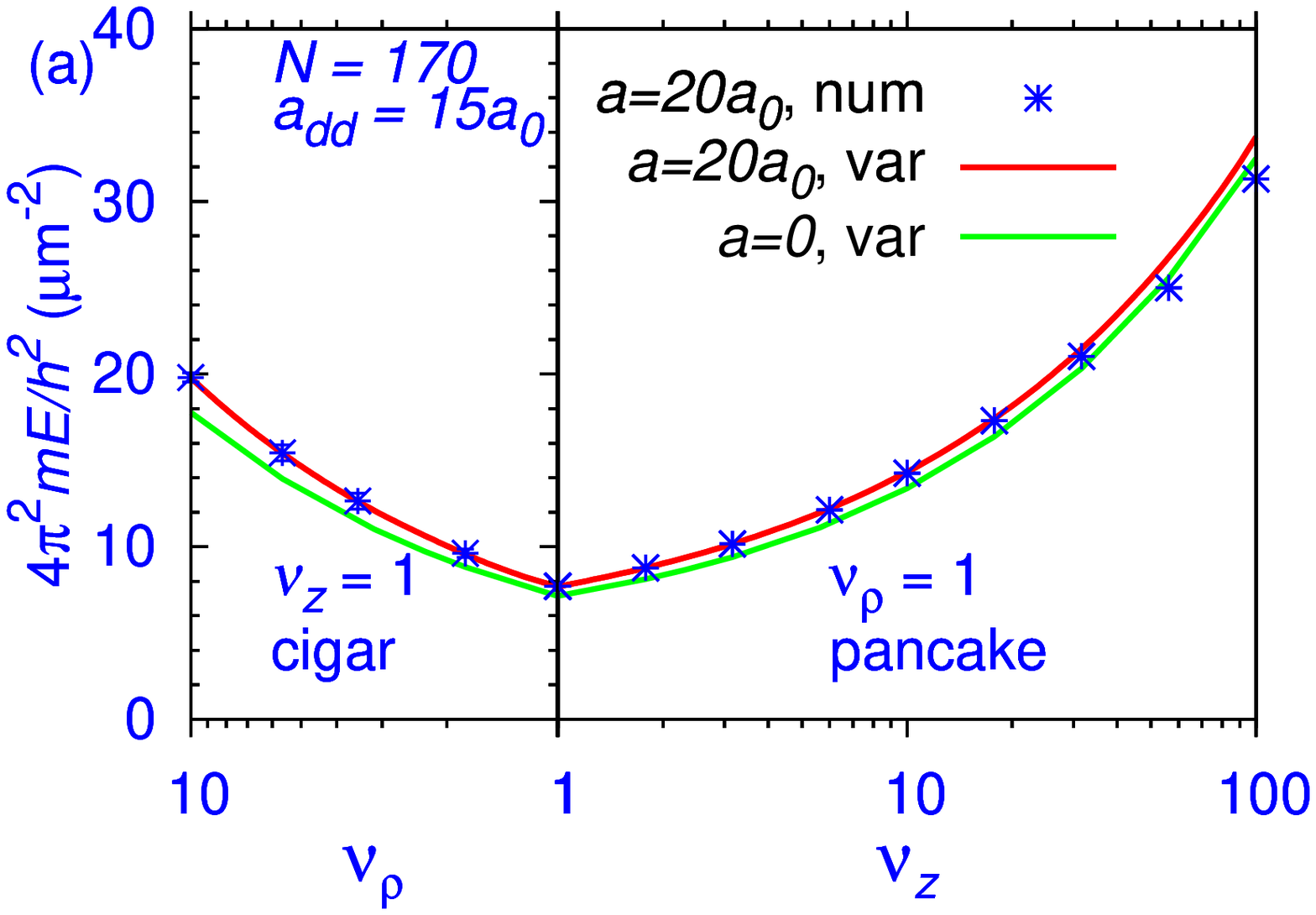}
\includegraphics[width=.49\linewidth]{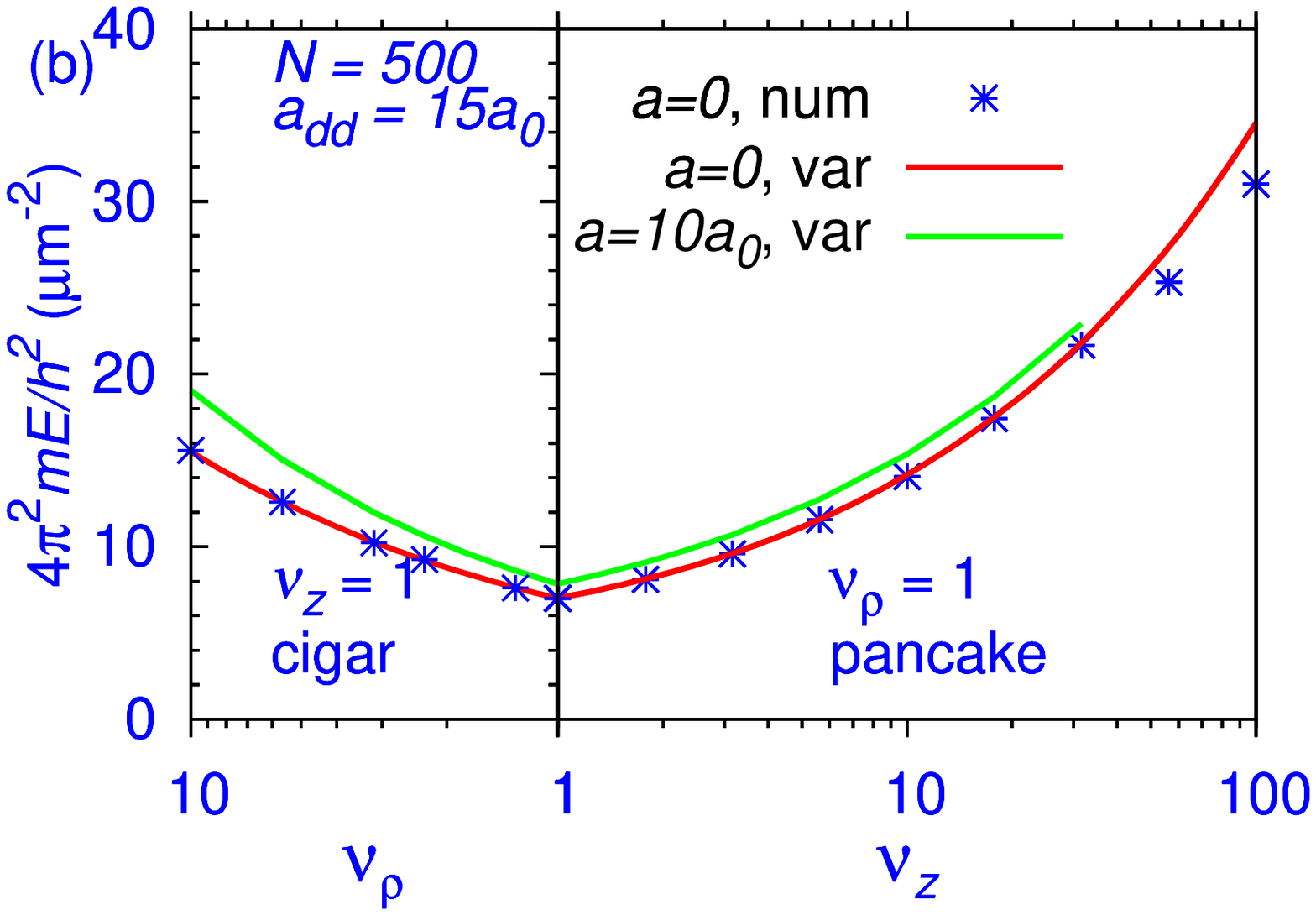}
\end{center}

\caption{(Color online) Numerical (num) and
variational (var) energy $E$ versus relative strengths of bichromatic OL 
$\nu_\rho$ and $\nu_z$ for (a)
$170$ and (b) $500$ $^{52}$Cr atoms.  }
\label{fig3} \end{figure}

From  (\ref{kappa}) we find that $f(\kappa)$ is positive for
cigar shape leading to an attractive contribution to
energy (\ref{enmin}), whereas it is negative in the pancake
shape leading to a repulsive term in energy. For $N=170$ there is a
symmetric state with $w_z=w_\rho$ and $f(\kappa)=0$
near $\nu_\rho=1, \nu_z \approx 1.3$ where the dipolar interaction
does not contribute and where the DBEC is most stable at the
 maxima in  figure
 \ref{Fig2} (b).
At this point the DBEC acts like a ``normal BEC" and the
$Na/a_0$ value at the maximum is independent of $N$.
In the pancake shape,
the localization of the  DBEC can be easily destroyed
 due to a large repulsive
dipolar interaction in the weak bichromatic OL. In the cigar shape,
the large attractive dipolar interaction may lead to the collapse of the
localized DBEC more easily than in the absence of dipolar interaction.
These aspects are clearly illustrated in  figure
 \ref{Fig2} (b). For $N=500$
the interactions are  stronger than for $N=170$ and the domain of 
allowed localization in terms of number of atoms
has reduced.

\begin{figure}
\begin{center}
\includegraphics[width=.45\linewidth]{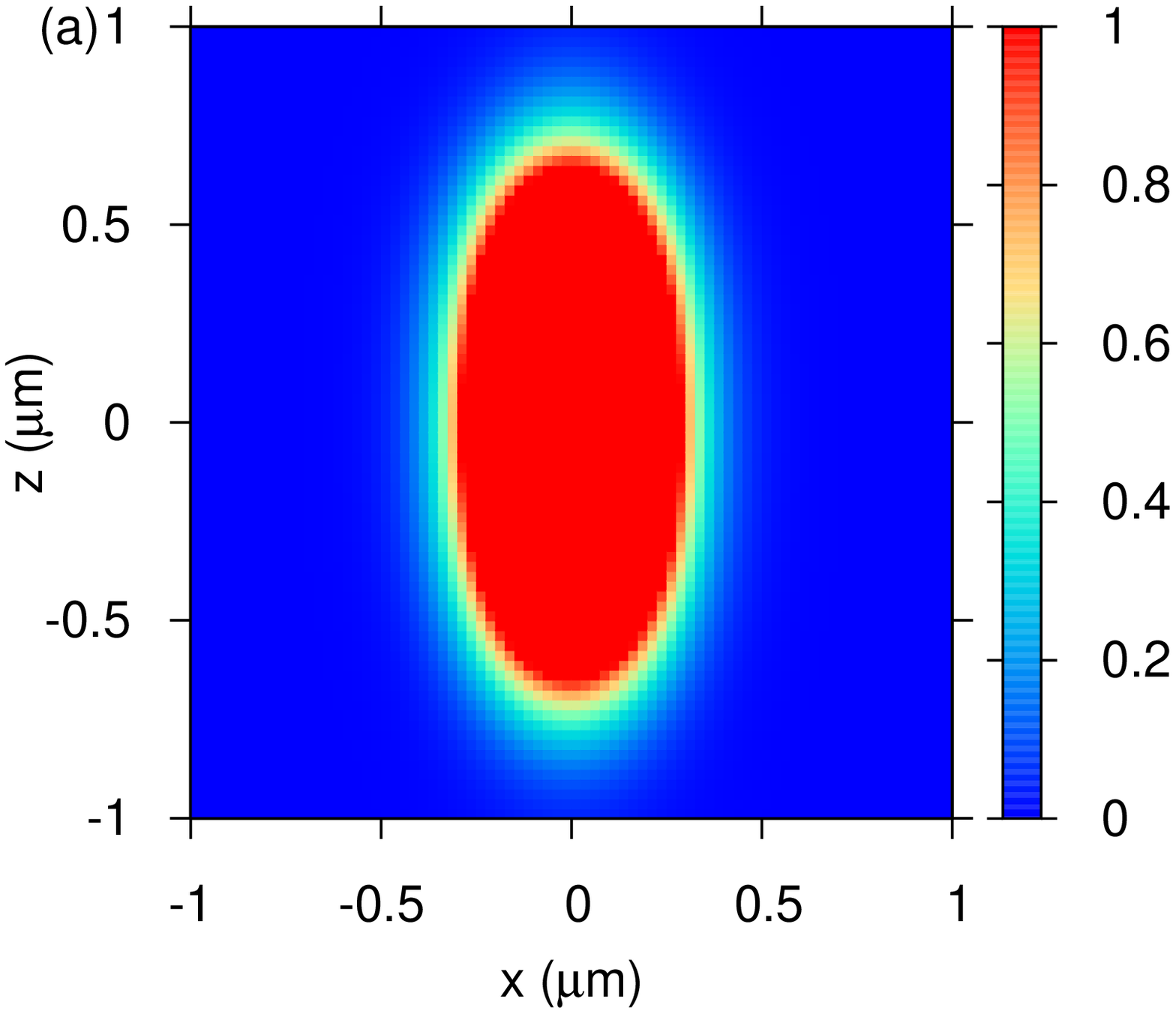}
\includegraphics[width=.45\linewidth]{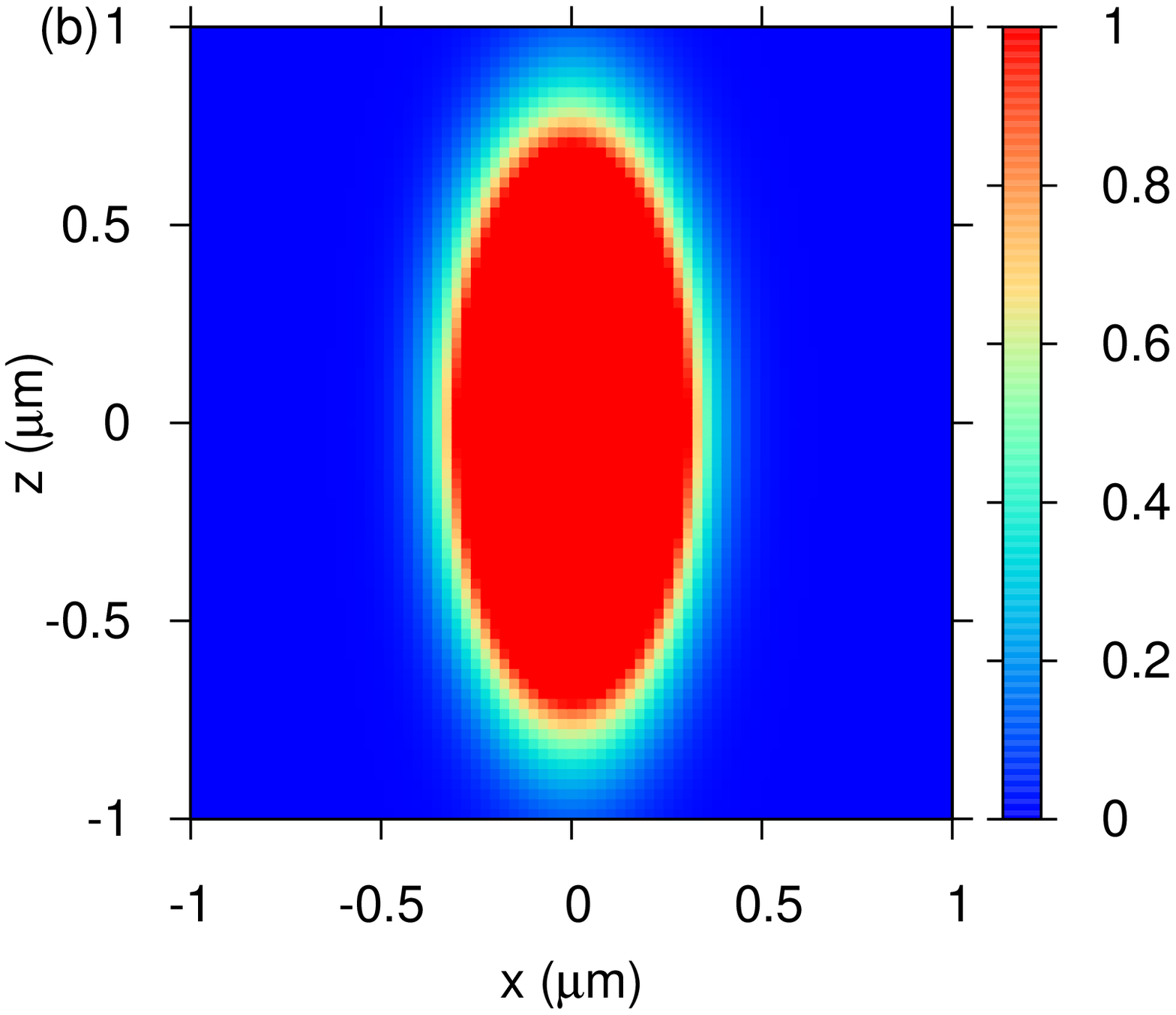}
\includegraphics[width=.45\linewidth]{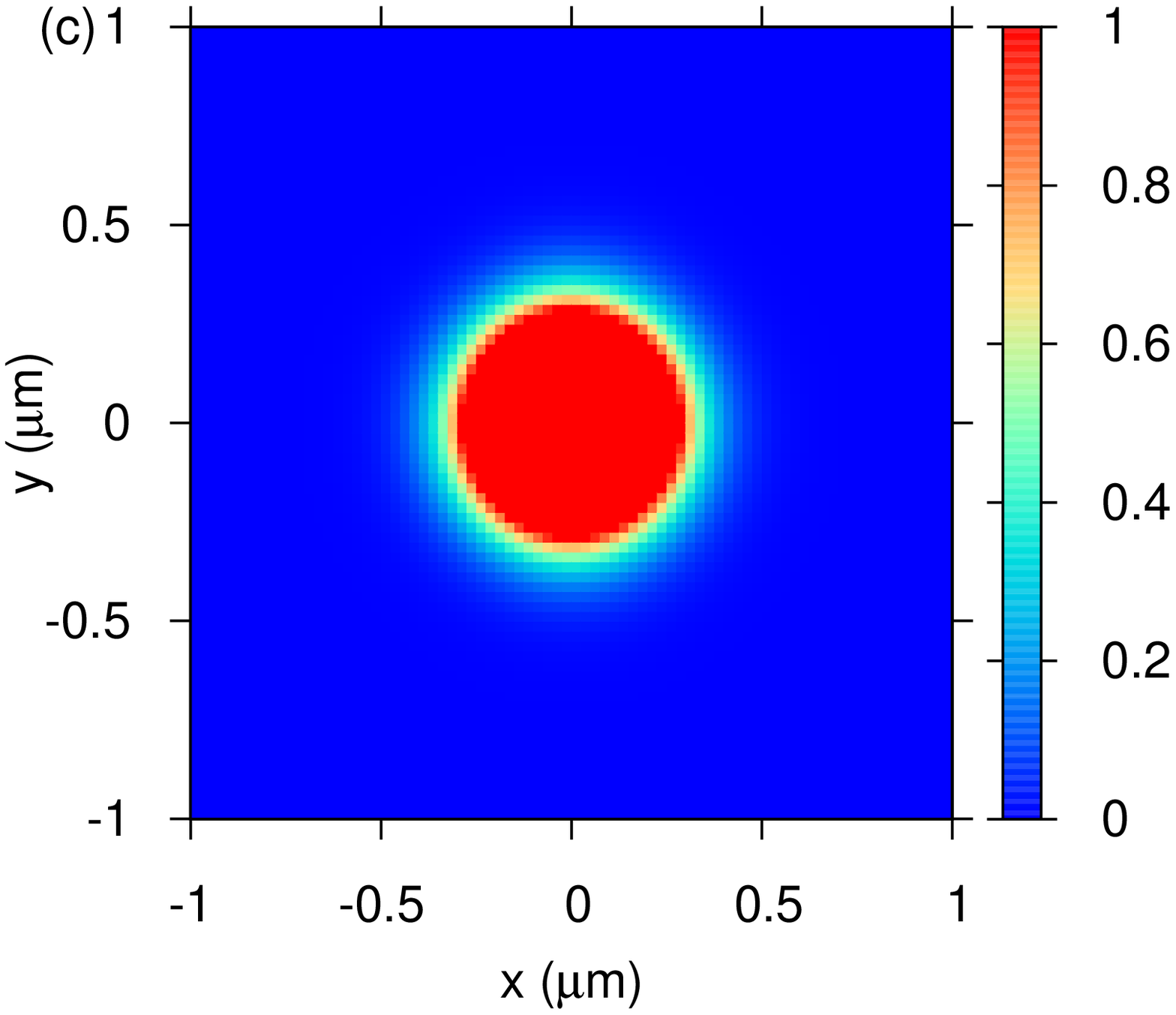}
\includegraphics[width=.45\linewidth]{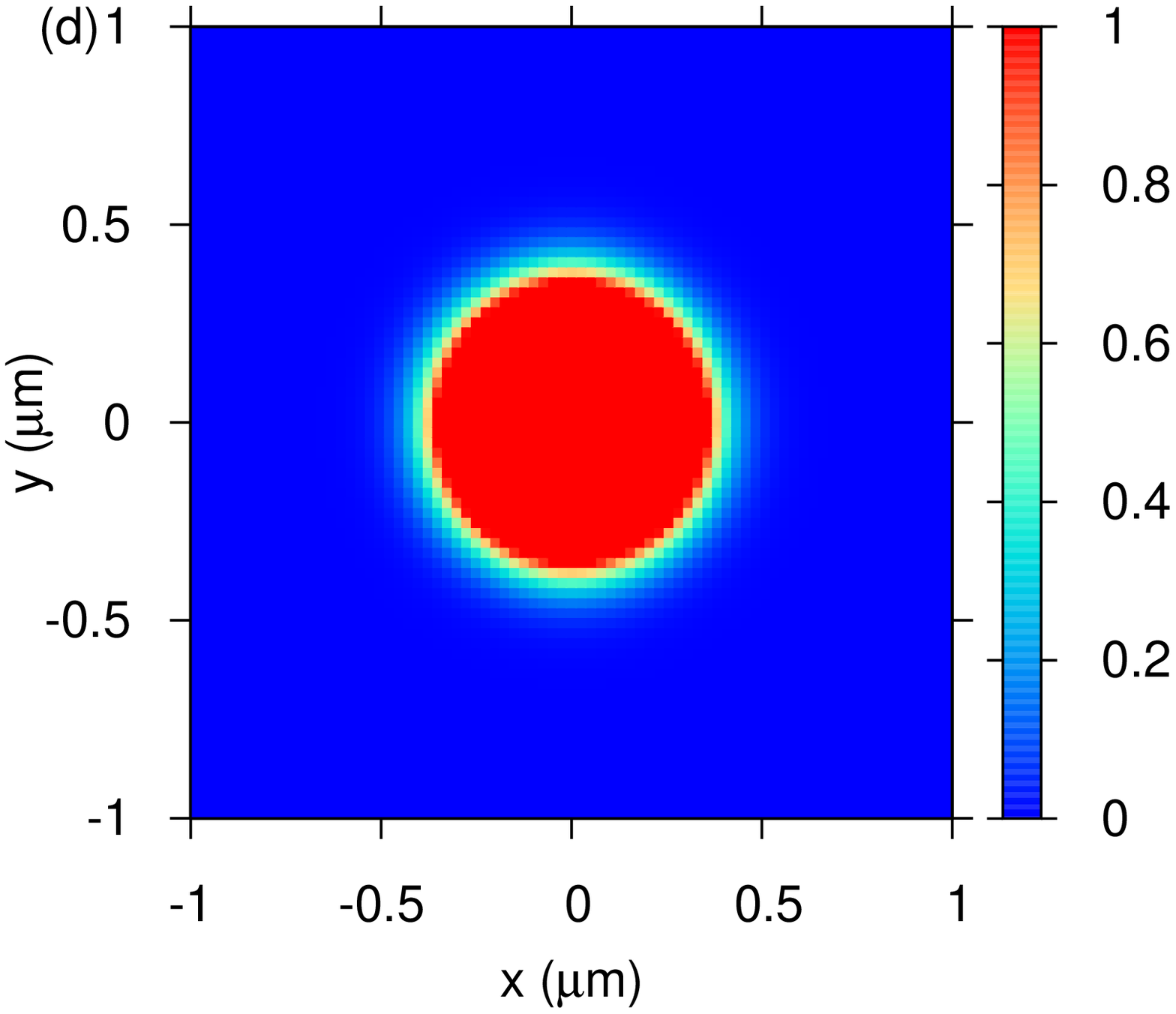}
\end{center}

\caption{(Color online)
2D Contour plot of density $\phi^2(x,0,z)$
in the $y=0$ plane for the cigar-shaped DBEC with $\nu_\rho=10, \nu_z=1$
of $N=500$ $^{52}$Cr atoms with $a=0$:
(a) Numerical  (b) variational.
2D Contour plot of density $\phi^2(x,y,0)$
in the $z=0$ plane for the same DBEC:
(c) Numerical  (d) variational.
} \label{fig4} \end{figure}

In  figures
 \ref{fig3}
(a) and (b) we compare the numerical and variational energies of the
$^{52}$Cr DBEC of 170 atoms with $a=0$ and $20a_0$ and of 500 atoms with
$a=0$ and $10a_0$, respectively. A smaller value of $a$ is required for
the stability of 500 atoms [see  figure
 \ref{Fig2} (b)]. The agreement
between variational and numerical results is good in general except for
$\nu_z \to 100,
\nu_\rho =1$. In this limit the localized DBEC occupies
two sites of the bichromatic OL and does not have a Gaussian shape.  
This justifies
a small disagreement between the numerical and variational results in
this limit.
The variational result for
$a=10a_0$ in  figure
 \ref{fig3} (b) only exists up to $\nu_z\approx 35$
[see  figure
 \ref{Fig2} (b)].

\begin{figure}
\begin{center}
\includegraphics[width=.45\linewidth]{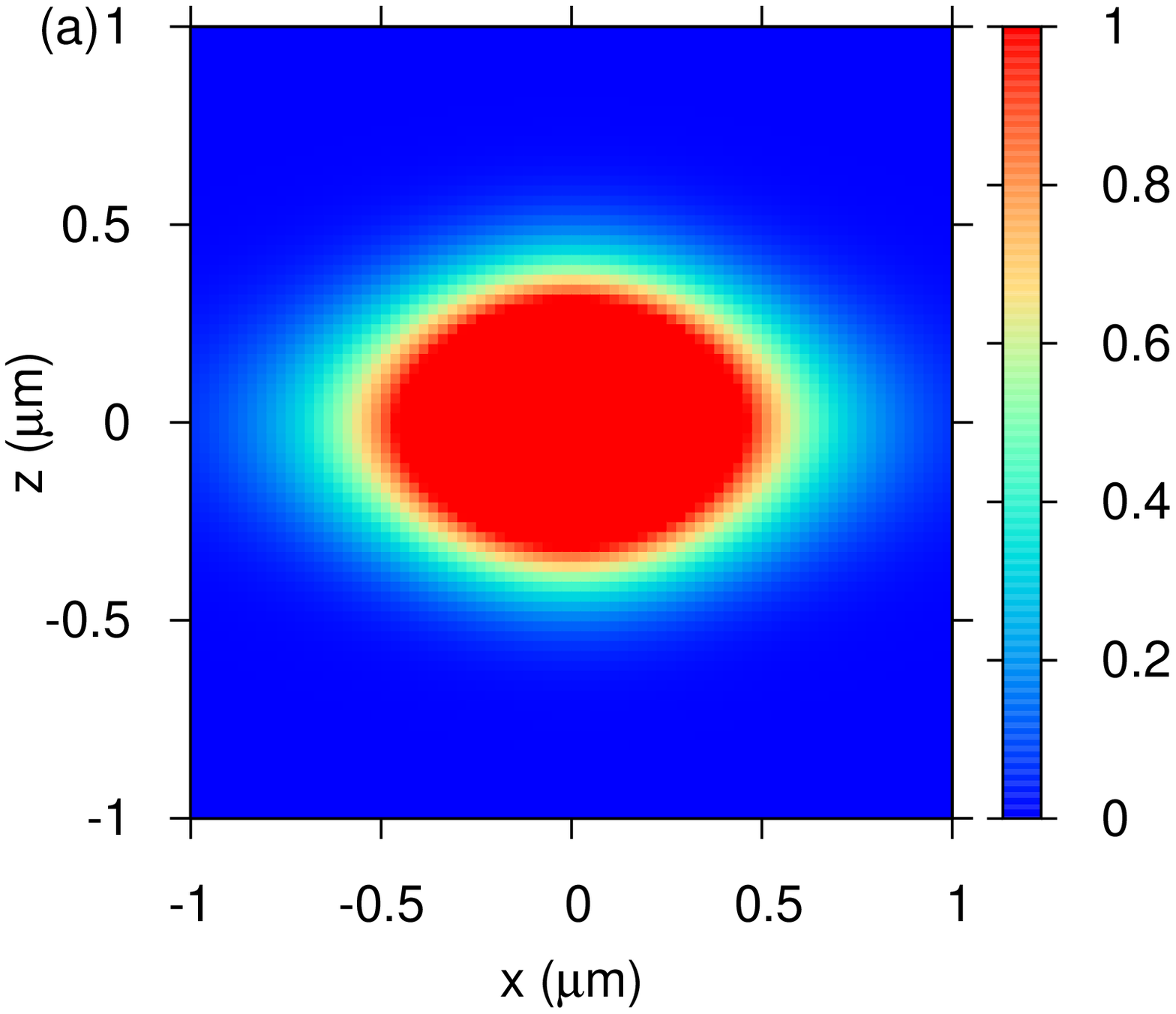}
\includegraphics[width=.45\linewidth]{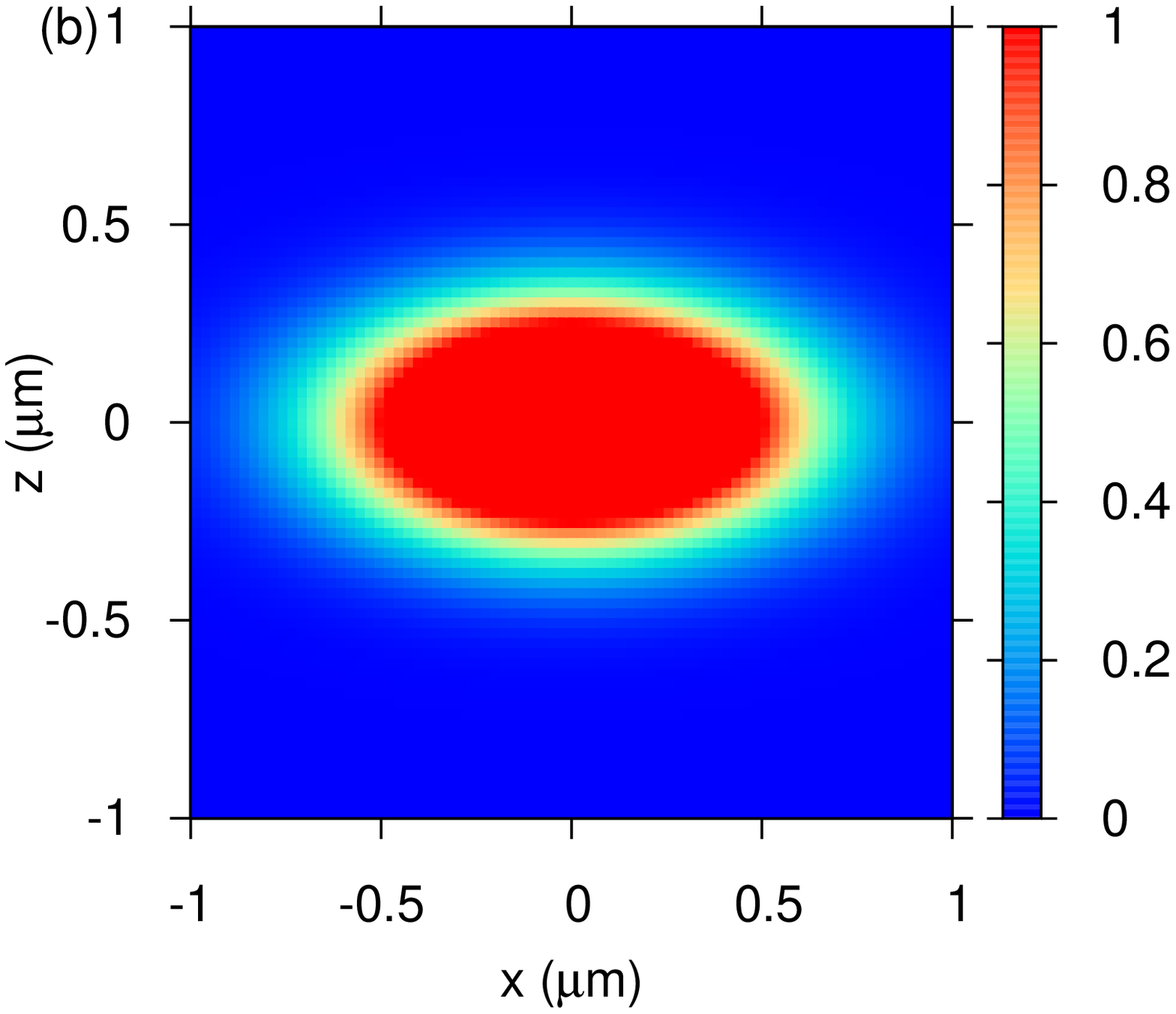}
\includegraphics[width=.45\linewidth]{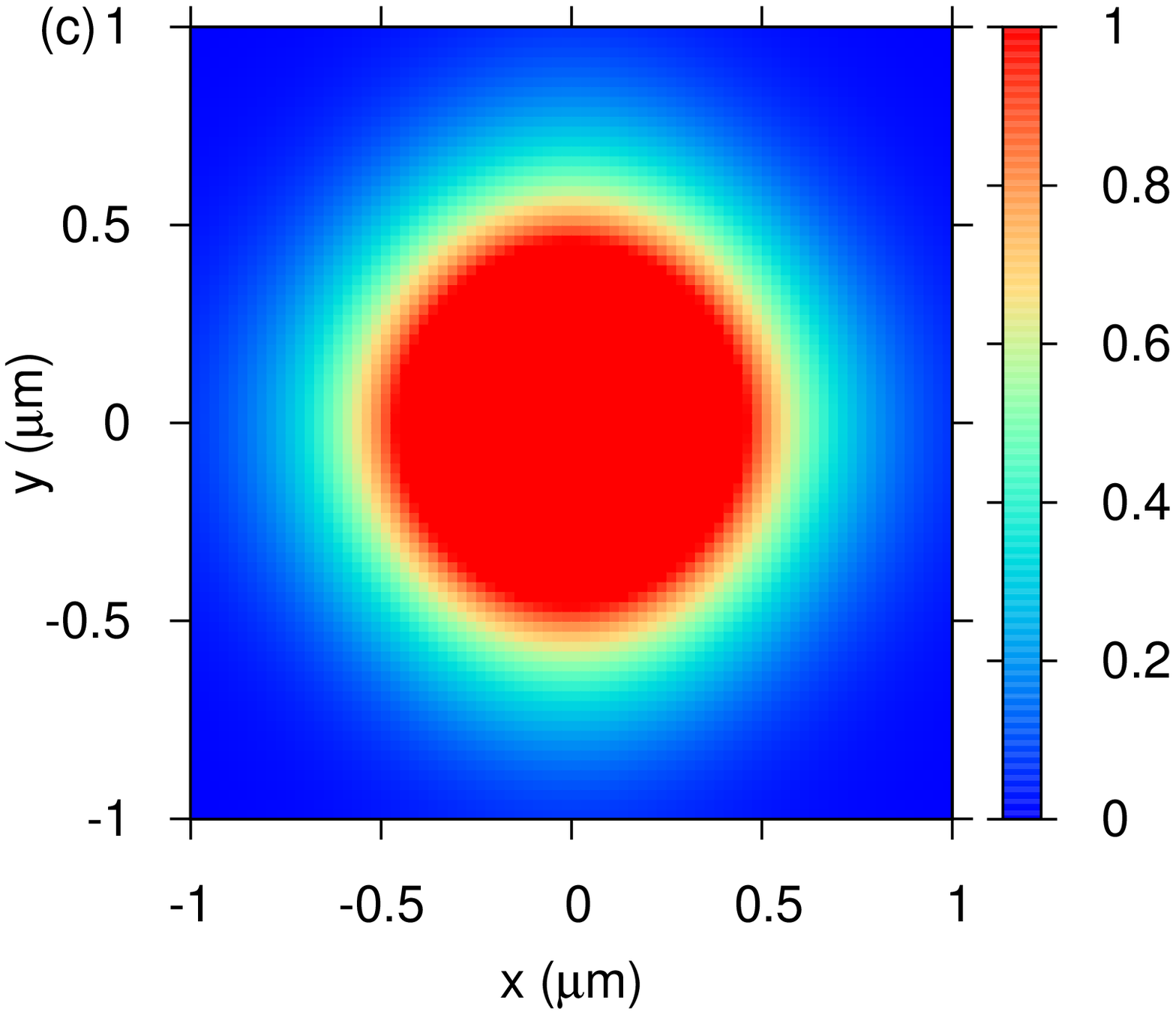}
\includegraphics[width=.45\linewidth]{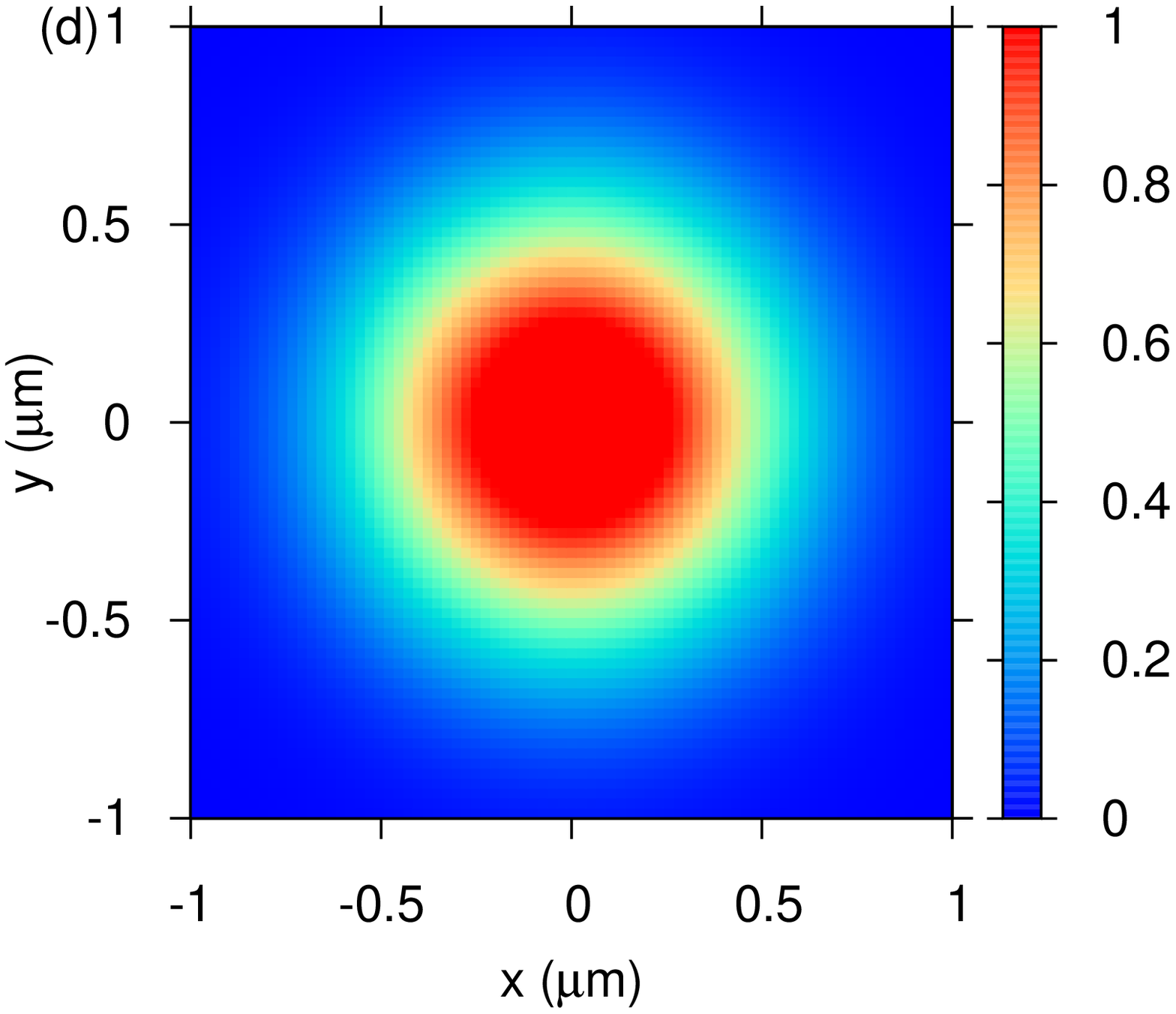}
\end{center}

\caption{(Color online)
Same as in  figure
 \ref{fig4} for the DBEC in the
pancake trap with $\nu_\rho=1, \nu_z=10$.
}
\label{fig5} \end{figure}

Next we study how the DBEC solely under the effect of dipolar
 interaction ($a=0$)
changes its shape as we move from the cigar to pancake shaped
configuration.
In figures \ref{fig4} (a) and (b)  we show the 2D contour plot from
the numerical and variational analysis, respectively, for density
$\phi^2(x,0,z)$ in the $y=0$ plane for a cigar-shaped DBEC
with $\nu_\rho=10, \nu_z=1$ for 500 $^{52}$Cr atoms for $a=0$. The 2D
contour plot from the numerical and variational
analysis, respectively,  for
density
$\phi^2(x,y,0)$ in the $z=0$ plane for
the same DBEC is shown in  figures \ref{fig4} (c) and (d), respectively.
This  localized numerical DBEC state  is of small size, compared to its
variational  counterpart,
due to the attractive dipolar interaction
in the cigar-shaped DBEC. The attraction due to dipolar interaction
has caused the DBEC to
contract
from an average Gaussian shape.
The corresponding numerical and
variational energies of the localized states of figures \ref{fig4}
are 15.16  and 15.53, respectively.
The localized  DBEC states in figures \ref{fig4}  (and also in  figures \ref{fig5})
occupy practically a single bichromatic OL site at the origin 
along $x,y$ and $z$:
$-1.25$ $\mu$m $ < x,y,z < 1.25$ $\mu$m. However, these localized states have
exponential tails.

The effect of the dipolar interaction is small  in the symmetric case with
$\nu_z=\nu_\rho=1$ and $N=500$.  Nevertheless, due to the complicated  
dipolar interaction,
the density $\phi^2(x,0,z)$ does not have a pure circular shape in this case
(not explicitly shown).
This density distribution
will be  circular for a BEC without dipolar interaction.
(A nearly circular shape for this density is obtained for a DBEC with
$\nu_\rho=1$ and $\nu_z=1.6$.)
The numerical and variational energies
for this state are  6.97 and 7.06, respectively.

We next consider the localization for $\nu_z=10, \nu_\rho=1$ and $a=0$.
In this pancake shape, the dipolar interaction is repulsive and for $500$ $^{52}$Cr
atoms a localized DBEC is obtained. In figures \ref{fig5} (a) and (b) we
show the numerical and variational 2D contour plots of density
$\phi^2(x,0,z)$ in the $y=0$ plane, respectively, for this DBEC. The
numerical and variational 2D contour plots for density $\phi^2(x,y,0)$
in the $z=0$ plane for the same DBEC is shown in figures \ref{fig5} (c)
and (d), respectively. In this case, the dipolar interaction is repulsive in nature and
hence the size of the DBEC is larger than the Gaussian variational shape
(see figures \ref{fig4} where the opposite happens for a DBEC with
attractive dipolar interaction.). The numerical and variational energies for this state
are 14.16 and 14.26, respectively. The change from a cigar shape to a
pancake shape of the localized DBEC as we move from $\nu_\rho = 10$ and
$\nu_z=1$ to $\nu_\rho = 1$ and $\nu_z=10$ is obvious from the density
distribution in figures \ref{fig4} (a) and \ref{fig5} (a).  Although the
variational and numerical energies (studied in figures \ref{fig4} and
\ref{fig5}) are quite close to each other, the numerically obtained
matter density should have some peculiarities not obtainable from the
variational calculation due to the anisotropic dipolar interaction. (The variational
calculation is based on an axially-symmetric Gaussian distribution.)

\section{Summary and Conclusion}
 We investigated the  localization 
of a $^{52}$Cr DBEC with in a weak bichromatic OL trap in the
presence and absence of short-range interaction using the numerical and 
variational solution
of the 3D GP equation (\ref{gp3d}). Of the two solutions, 
the numerical solution is the most precise
one and should be used  in case of disagreement with the variational solution.
   Although the density of the central
part of the localized states has a near Gaussian distribution, the
density distribution also has a long exponential tail 
 \cite{billy,roati}. The
Gaussian distribution near the center permits a
variational analysis of localization, which is used for an analytical
understanding of the problem.
A DBEC of a small number of atoms with a weak
short-range interaction could be
localized by a relatively weak bichromatic OL trap.  From the  variational
solution  we obtain  a phase diagram [figure
 \ref{Fig2} (b)] illustrating the effect of the dipolar interaction on 
the localization as a function of the strengths of the trap $\nu_z$ and 
$\nu_\rho$ in axial and radial directions, respectively. We find that 
for $^{52}$Cr atoms, the dipolar interaction has a moderate effect 
on localization. (Larger effect will certainly appear for dipolar 
molecules where the dipole moment could be larger by an order of 
magnitude compared to the dipole moment of $^{52}$Cr atoms.) The 
numerical and variational energies of the DBEC, as well as the 
corresponding densities are in reasonable
 agreement with each other. In the absence of a short-range interaction, 
the localized DBEC can accommodate the largest number of $^{52}$Cr atoms 
($\sim$ 1000) in the spherical configuration and this number reduces 
both for cigar and pancake shapes due to the attractive and repulsive 
dipolar interaction, respectively. The attractive dipolar 
interaction leads to collapse and the repulsive dipolar 
interaction leads to leakage to infinity. We hope that this study will 
motivate experiments on the localization of a $^{52}$Cr DBEC in a 
bichromatic OL trap.  The estimate of the number of localized $^{52}$Cr 
atoms, their radial and axial sizes and shapes etc. as predicted in the 
present study can be verified in the experiment.

\ack

FAPESP and  CNPq (Brazil) and DST (India)
provided partial support.

\vskip 0.5 cm
References


\vskip 0.5 cm

\begin{thebibliography}{99}

\bibitem{anderson}  Anderson P W 1958 {\it Phys. Rev.} {\bf 109}  1492 


\bibitem{chabe}  Chab\'e  J {\it et al.} 2008  {\it Phys. Rev. Lett.} {\bf 101}
255702 

 Edwards E E {\it et al.} 2008 \PRL 
 {\bf 101} 260402 


\bibitem{billy}  Billy  J  {\it et al.} 2008 {\it Nature } {\bf 453} 891 

\bibitem{roati}  Roati G {\it et al.} 2008  {\it Nature }  {\bf 453} 895 

\bibitem{fesh}  Inouye  S  {\it et al.} 1998
{\it Nature} 1998  {\bf 392} 151 



\bibitem{aubry}  Aubry S  and  Andr\'e G  1980 {\it Ann. Israel  Phys. Soc.}
{\bf 3} 133 

  Kopidakis G,  Komineas S,  Flach S and  Aubry S 2008 
{\it Phys. Rev. Lett.} {\bf 100} 084103 

\bibitem{Albert}  Albert M and  Leboeuf P  2010 {\it Phys. Rev. A} {\bf 81} 013614








\bibitem{boers} Boers D J,  Goedeke B,  Hinrichs D and
Holthaus M 2007  {\it Phys. Rev. A} {\bf 75} 063404

 Sanchez-Palencia  L {\it et al.} 2007 {\it  Phys. Rev. Lett.}
{\bf 98} 210401 

  Cl\'ement D {\it et al.} 2005  {\PRL}
{\bf 95} 170409 

 Lye  J E {\it et al.}  2005  \PRL
{\bf 95} 070401 

  Damski B {\it et al.} 2003 \PRL
{\bf 91} 080403 

 Schulte T {\it et al.}   2005  \PRL
{\bf 95} 170411 

 Roux  G {\it et al.,} 2008 
{\it Phys. Rev. A} {\bf 78} 023628 

 Roscilde T 2008 {\it Phys. Rev. A} {\bf 77} 063605

  Paul T {\it et al.}  {\it Phys. Rev. A} {\bf 80} 033615



\bibitem{modugno}
 Modugno M  2009 {\it New. J. Phys.} {\bf 11} 033023 

  Larcher M,  Dalfovo F and  Modugno M 2009 
{\it Phys. Rev. A} {\bf 80} 053606 


\bibitem{adhikari}


Cheng Y  and  Adhikari S K 2010 
{\it Phys. Rev. A}
 {\bf 82} 013631  

 Adhikari S K and  Salasnich L 2009 
{\PR A} {\bf 80} 023606 

\bibitem{cheng}

Cheng Y  and  Adhikari S K 2010 
{\it Phys. Rev. A}
 {\bf 81} 023620  


\bibitem{adhikari1}
 Adhikari S K 2010  {\it Phys. Rev. A} {\bf 81} 043636


\bibitem{pfau} Lahaye T {\it et al.} 2007  {\it Nature} {\bf 448} 672

Lahaye T {\it et al.} 2009  {\it Rep. Prog. Phys.}
{\bf 72} 126401 

 Koch T {\it et al.} 2008 {\it Nature Phys.} {\bf 4} 218

 Stuhler  J {\it et al.} 2005 {\it  Phys. Rev. Lett.} {\bf 95} 150406 

 Griesmaier A {\it et al.} 2005  \PRL 
{\bf 94} 160401 


\bibitem{jb}
 G\'oral K and  Santos L 2002  {\it Phys. Rev. A} {\bf 66} 023613 

  Yi S and  You L 2001  {\it Phys. Rev. A}
{\bf 63} 053607 



\bibitem{Dutta2007}
 Dutta O and  Meystre P 2007 {\it Phys. Rev. A}
 {\bf 75} 053604 

\bibitem{Parker2009}Parker N G, Ticknor C, Martin A M and 
 O'Dell D H J 2009 {\it Phys.
  Rev. A} {\bf 79}  013617

Ticknor C,  Parker N G, Melatos A, Cornish S L, O'Dell D H J and 
  Martin A M
2008 {\it  Phys. Rev. A}
{\bf   78 } 061607



\bibitem{other}
 Santos L,  Shlyapnikov G V and  Lewenstein M 2003 {\it  Phys. Rev.
Lett.} {\bf 90} 250403 

  Yi S,  You L 
  and  Pu H 2004 {\it Phys. Rev. Lett.} {\bf 93} 040403 

 G\'oral K,  Santos L and  Lewenstein M 2002 {\it Phys. Rev. Lett.}
{\bf 88} 170406 

 Giovanazzi S, O'Dell D and  Kurizki G 2002 {\it  Phys. Rev.
Lett.} {\bf  88} 130402 



\bibitem{metz} Metz  J {\it et al.} 2009 {\it 
 New J. Phys.} {\bf 11} 055032 


\bibitem{coll}
 Sun B and  Pindzola M S 2009 {\it  J. Phys. B} {\bf 42}
175301 

  Lahaye  T {\it et al.} 2008 {\it Phys. Rev. Lett.} {\bf 101}
080401 




\bibitem{Ronen2006a} Ronen S, Bortolotti D C E and Bohn J L 2006
{\it Phys. Rev. A}
  {\bf 74} 013623





\bibitem{Yi2000}
Yi S and  You L 2000 {\it Phys. Rev. A}
  {\bf 61} 041604

\bibitem{Yi2003}
Yi S and  You L 2003 {\it Phys. Rev. A}
  {\bf 67} 045601


\bibitem{Santos2000} Santos L, Shylapnikov G V, Zoller P and  Lewenstein M
2000  {\it Phys. Rev. Lett.}  {\bf 85} 1791


\bibitem{Wilson2008}
 Wilson R M, Ronen S,  Bohn J L  and Pu H 2008 {\it Phys. Rev. Lett.} 
{\bf 100}  245302


\bibitem{Ronen2007}
Ronen S, Bortolotti D C E and Bohn J L 2007
{\it Phys. Rev. Lett.} {\bf  98}  030406


\bibitem{Wilson2009}
 Wilson R M, Ronen S and  Bohn J L  2009
 {\it Phys. Rev. A} {\bf 80} 023614



\bibitem{2D3D}  Kuhn R C {\it et al.} 2005 {\it  Phys. Rev. Lett.} {\bf 95}
250403 

   Skipetrov S E {\it et al.} 2008 {\it  Phys. Rev. Lett.}
 {\bf 100}
165301 



\bibitem{destruction}  Pikovsky A S and  Shepelyansky D L 2008 
{\it Phys. Rev.
Lett.} {\bf 100} 094101 

 Flach S,  Krimer D O  and  Skokos Ch 2009 
{\it Phys. Rev. Lett.} {\bf 102} 024101 








\bibitem{you} Yi S and  You L 2004 {\it Phys. Rev. Lett.} {\bf 92} 193201 

\bibitem{murug}Adhikari S K and   Muruganandam P 2003 \jpb 
{\bf 36} 409

 Muruganandam P and Adhikari S K 2003 \jpb
{\bf 36} 2501



\bibitem{CPC}  Muruganandam P and Adhikari S K 2009 {\it Comput. Phys. Commun.}
{\bf 180}
 1888 



\bibitem{self} Smerzi A,  Fantoni S,  Giovanazzi S and  Shenoy S R
1997 
{\it Phys. Rev. Lett.} {\bf 79} 4950 

  Raghavan S, 
Smerzi A,  Fantoni S  and  Shenoy S R 1999  {\it Phys. Rev. A} {\bf 59}
620 

 Albiez M,  Gati R,  Folling J,  Hunsmann S,  Cristiani M
and  Oberthaler M K 2005 {\it  Phys. Rev. Lett.} {\bf 95} 010402


  Gati R and  Oberthaler M K 2007 {\it  J. Phys. B} {\bf 40}
R61


  Adhikari S K,  Lu H and  Pu H  2009 {\it Phys. Rev. A} {\bf 80}
063607 








 \end{thebibliography}
\end{document}